\documentclass[sigconf]{acmart}

\def\BibTeX{{\rm B\kern-.05em{\sc i\kern-.025em b}\kern-.08emT\kern-.1667em\lower.7ex\hbox{E}\kern-.125emX}}
    
\setcopyright{none}
\renewcommand\footnotetextcopyrightpermission[1]{} 
\hypersetup{draft}

\usepackage{booktabs} 
\usepackage{graphicx}
\usepackage{subcaption}
\usepackage{balance}

\usepackage{listings}
\usepackage{xcolor}

\colorlet{punct}{red!60!black}
\definecolor{background}{HTML}{EEEEEE}
\definecolor{delim}{RGB}{20,105,176}
\colorlet{numb}{magenta!60!black}

\lstdefinelanguage{json}{
    basicstyle=\normalfont\scriptsize,
    numbers=left,
    numberstyle=\scriptsize,
    stepnumber=1,
    numbersep=8pt,
    showstringspaces=false,
    breaklines=true,
    frame=lines,
    backgroundcolor=\color{background},
    literate=
     *{0}{{{\color{numb}0}}}{1}
      {1}{{{\color{numb}1}}}{1}
      {2}{{{\color{numb}2}}}{1}
      {3}{{{\color{numb}3}}}{1}
      {4}{{{\color{numb}4}}}{1}
      {5}{{{\color{numb}5}}}{1}
      {6}{{{\color{numb}6}}}{1}
      {7}{{{\color{numb}7}}}{1}
      {8}{{{\color{numb}8}}}{1}
      {9}{{{\color{numb}9}}}{1}
      {:}{{{\color{punct}{:}}}}{1}
      {,}{{{\color{punct}{,}}}}{1}
      {\{}{{{\color{delim}{\{}}}}{1}
      {\}}{{{\color{delim}{\}}}}}{1}
      {[}{{{\color{delim}{[}}}}{1}
      {]}{{{\color{delim}{]}}}}{1},
}

\begin{document}

\title{A Semantic Schema for Data Quality Management in a Multi-Tenant Data Platform}

\renewcommand{\shorttitle}{A Semantic Schema for Data Quality Management}

\author{Ning Zhou}
\affiliation{%
  \institution{Microsoft}
  \city{Oslo}
  \country{Norway}
}
\email{ning@duoja.com}
\authornote{\label{authornote}The work described in this paper was done when the author worked at Schibsted Media Group.}

\author{Sandra Garcia Esparza}
\affiliation{%
  \institution{Schibsted Media Group}
  \city{Oslo}
  \country{Norway}
}
\email{sandra.garcia@schibsted.com}

\author{Lars Marius Garshol}
\affiliation{%
  \institution{Schibsted Media Group}
  \city{Oslo}
  \country{Norway}
}
\email{larsga@garshol.priv.no}

\renewcommand{\shortauthors}{N. Zhou, S. Garcia Esparza and L. M. Garshol}

\begin{abstract}
Schibsted Media Group is a global marketplace company with presence in more than 20 countries. It is undergoing a digital transformation to convert data silos to a multi-tenant system based on a common data platform. Good data quality based on a common schema on the semantic level is essential for building successful data-driven products across marketplaces. To solve this challenge, we developed the data quality tooling based on a semantic schema management system to support schema evolution with versioning, testing and transformation. It can monitor the data quality requirements for different applications and handle incoming data consisting of multiple schema versions. Today the system is operating in production and processes over one billion events per day for over 100 applications.

\end{abstract}

%
%



\maketitle

\section{Introduction}
\label{sec:intro}

\begin{figure}
\centering
\includegraphics[width=0.9\linewidth]{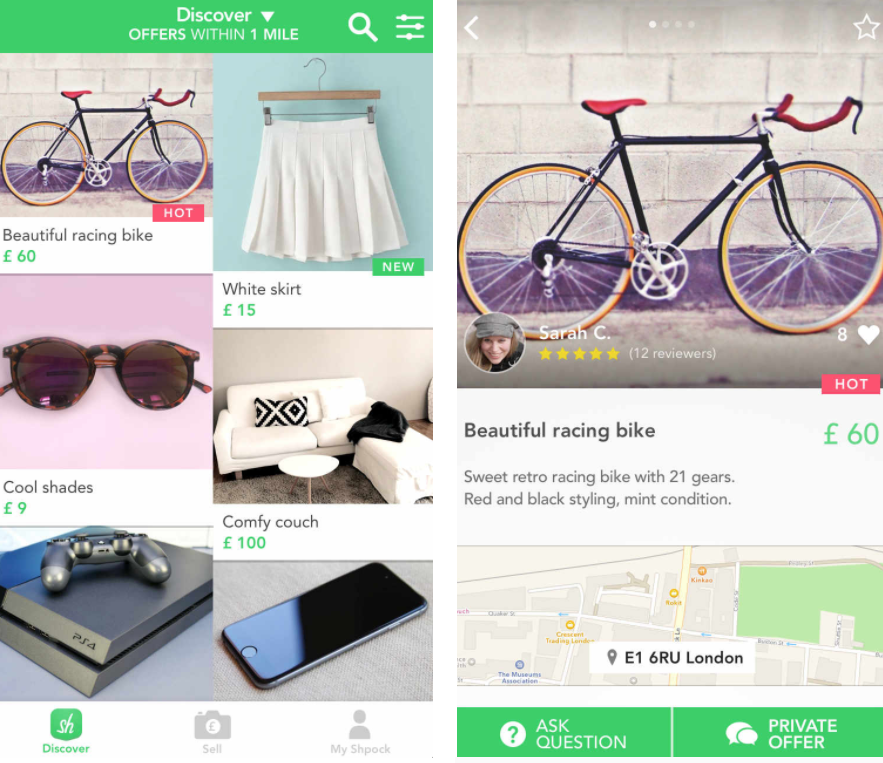}
\caption{Sample item feed (left) and item detail screen (right) from Shpock, a Schibsted marketplace.}
\label{fig:marketplace}
\end{figure}

Schibsted Media Group is a global online marketplace company that operates multiple verticals in more than 20 countries. Marketplaces provide a platform where sellers post their items for sale and buyers search for the items they want. The items can range from low-value ones such as books and clothes to high-value ones such as cars and real estate. Users can also post non-tangible products and services such as discount codes and job openings. To improve the user experience, marketplaces are often divided into several vertical markets adapted to the specific needs of a certain category. There are single-vertical marketplaces specialized in real state, car or jobs, as well as general marketplaces that have a rich variety of categories such as clothes, books and furniture similar to a regular e-commerce platform. An example of a Schibsted-owned general marketplace called Shpock (\url{https://en.shpock.com}) is shown in Figure \ref{fig:marketplace}. 

Marketplace users need search and discovery capabilities to navigate the vast amount of items posted for sale as well as smart anti-fraud measures to build trustworthy communities. Today's fast technology advances can help satisfy these user needs, so Schibsted initiated a digital transformation in 2014 to build data-driven innovations that can be shared by all its marketplaces. The company plans to move its polymorphic products to a multi-tenant system to consolidate most of the machine learning and infrastructure services to a common data platform for better cost efficiency while keeping the flexibility of per-market product localization.  

To materialize the synergies from the consolidation, processing of data across tenants is a must in order to make machine learning efforts scalable. However, the cost of migrating data silos of each marketplace onto a common data platform is not trivial. Building a multi-tenant data lake imposes not only technical but also organizational challenges. For instance, different tenants do not share the same schema, and properties with the same name can have different meanings for different tenants. Data producers and data consumers are often not in the same department of the company, and the communication cost between them are non-negligible. Thus we devised a semantic schema that defines both the syntax of the schema and the semantic meaning of each property name and value. It serves as a protocol to make sure that data are ingested in a consistent way from different tenants across the company. In this way, data consumers can not only parse and access data types and values correctly, but also interpret the meanings of the values correctly.

There is no shortage of existing solutions to facilitate wrangling data from multiple sources. Commercial solutions such as Xiti (\url{https://www.xiti.com}) and Tealium (\url{http://tealium.com}) often use a simple list of key-value pairs to achieve schema-agnostic data collection. Such solutions define the syntax of the data, but cannot ensure the semantic correctness. For example, they can require a property to be a string value, but that does not guarantee the value is correct. They are flexible in the sense that new properties can be added easily. However, they do not detect data quality issues and instead push them downstream to data consumers. This is where having a semantic schema can help. Open-source communities try to crowd-source semantic schemas in a collaborative way. Prominent examples include Schema.org \cite{schema:online}, Activity Streams \cite{activitystreams:online} and Open Graph \cite{og:online}. We use them as a starting point and build upon them to cover the comprehensive needs of the specific use case at Schibsted.

In this paper, after a brief system overview in Section \ref{sec:overview}, we first describe the semantic schema system in Section \ref{sec:schema} and \ref{sec:tooling}, then the data quality tool and its usage in downstream machine learning applications in Section \ref{sec:quality}, and finally conclude the paper in Section \ref{sec:conclusion}. 

\begin{figure*}
\centering
\includegraphics[width=0.85\linewidth]{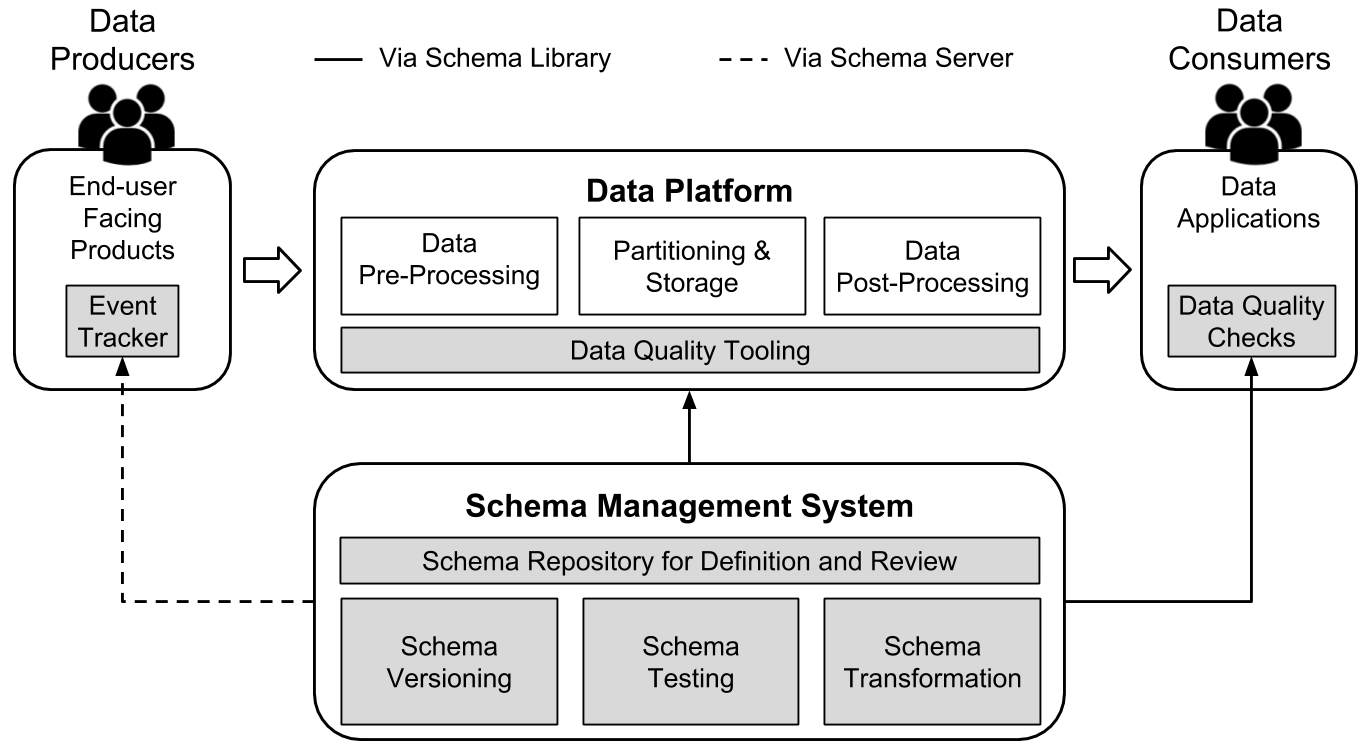}
\caption{The overview of the data quality management system in the end-to-end data solution. The components highlighted in gray are covered in details in this paper.}
\label{fig:system}
\end{figure*}

\section{Related work}
\label{sec:related}

A similar schema system to the one we introduced in this paper for standardizing and unifying events and entities in a multi-tenant system is presented in\cite{Sumbaly:2013} which describes the data pipeline used at LinkedIn. Whereas their paper focuses more on the the data pipeline and the applications they use on top of it, our emphasis is the schema management and the tooling to monitor data quality.

Some of the challenges we describe in this paper such as schema evolution have also been discussed in previous works. For example, in \cite{Curino08} the authors present a study that shows the evolution history of Wikipedia schemas, which are updated by a large number of groups and individuals. Their work targets at relational database schemas where schema operators correspond to table operations such as \texttt{CREATE TABLE} or \texttt{ADD COLUMN}. Instead, our schema corresponds to data that lives in a lambda architecture supporting various batch and real-time applications.

Our overall solution is designed in a schema agnostic way similar to \textit{Azure DocumentDB} \cite{microsoft:nonschema}, so that the data platform can process different datasets without the need of building schema dependent features. The data discovery problem can be challenging in this setting, for the data consumers have to find the specific datasets they need. Google solves it with \textit{Goods} \cite{halevy:google}, an intelligent data indexer that extracts certain properties such as schema from datasets. Instead we propose to use a semantic schema to unify the vocabulary used by different tenants. Downstream applications can discover and understand the data they need by search and parse the schema meta data.

The process of matching existing dataset to a new schema is somewhat similar to entity extraction from semi-structured data and knowledge base construction using the Internet as a source. The research in this area mainly focuses on how to use automation to reduce human efforts needed. Examples in the e-commerce related area include DEXTER \cite{qiu:dexter} and DIADEM \cite{furche:diadem} that parse e-commerce websites automatically to extract structured product data. There are also examples such as Wrangler \cite{kandel:chi} and user-involved data cleaning system \cite{galhardas:cleaning} that leverage human feedback to construct data transformations. In this paper, schema auto generation is not our focus due to the limited scope of the system - it is designed to solve the internal needs of a common data model for one company, which is more suitable for manual design and update. But automation, in particular auto transformation between different schema versions, is indeed an area that we plan to explore in future.

\section{System Overview}
\label{sec:overview}

The core of the multi-tenant solution at Schibsted is a \textit{data platform} that ingests data from different marketplaces and make them available to the central and local teams building data-driven applications. Examples of such applications include a messaging service among end users, a recommendation engine, or an fraud detector dealing with malicious users. The data platform uses the lambda architecture to support both hourly batch and real-time streaming applications.

As depicted in Figure \ref{fig:system}, data quality management is based on a stand-alone schema management system which is integrated end-to-end in the pipeline. The semantic schemas not only define the format of the data, but also impose constraints on the content of the data. In addition to the schema definitions, the schema system also provides tooling to handle schema updates via versioning, testing and transformation. Data quality tooling based on the semantic schema system allows us to standardize the incoming data from all marketplaces to a common schema and get them processed, stored and distributed to the downstream data consumers. An event tracker developed in-house uses the schema library to facilitate data collection from end-user facing products. Data quality tooling in the data platform is responsible for using the evolving semantic schemas to route and process the data that flows through the pipeline. Data quality check enables data consumers to communicate their requirements to producers with telemetry and monitoring. They can access the schema and its tooling from the schema server.

The goal of this solution is not only to power more data-driven applications, but also to lower the cost of data flowing through organization boundaries. The schema semantic plays an important role in facilitating communications as well as enabling automation in the process of collecting and processing data. The end-to-end solution has been running in production at Schibsted for three years. As of today, it ingests over roughly 1 billion events per day (in total about 2.5 terabytes) from over 100 tenants. The solution is essential for data teams in Schibsted to achieve good scalability with model training on multiple tenants \cite{vanderWeide:2017}.

\section{Schema definition}
\label{sec:schema}

The schemas we have defined focus on the user behaviors in the marketplaces. Their definitions are stored in a Github (\url{https://github.com}) repository and are evolved in an incremental manner, as new needs from data consumers arise. A schema committee consisting of members from different data teams review the new schema contributions and may decide on important system design issues. 

The schema files should be both human and machine readable to facilitate using the schema in different scenarios. For instance, when working with a new dataset for the first time, data analysts and data scientists need to understand the schema in order to analyze and use the relevant information. Whereas in regular production jobs, schemas are more often read by telemetry code to ensure the data quality. For this reason we choose to define the schema in JSON \cite{json:online} format, because it can satisfy both needs. We divide schemas into two large categories: object and event. An \textit{event schema} is a complete data model that can be used by a dataset, while an \textit{object schema} is a fragment model that can be referenced by another schema to facilitate schema alignment. In addition to schema reference, we also support schema extensibility to reuse event schema by inheritance. 

\subsection{Semantics and syntax}
\label{sec:semantics}

The primary organizational challenge of a common schema is to establish common vocabularies across the organization. We want to build a common schema to describe the common behaviors that users perform in all marketplaces. For example, behaviors such as a user posting an item for sale, viewing an item, or contacting a seller happen in all marketplaces. 

\begin{figure}
\centering
\includegraphics[width=0.4\textwidth]{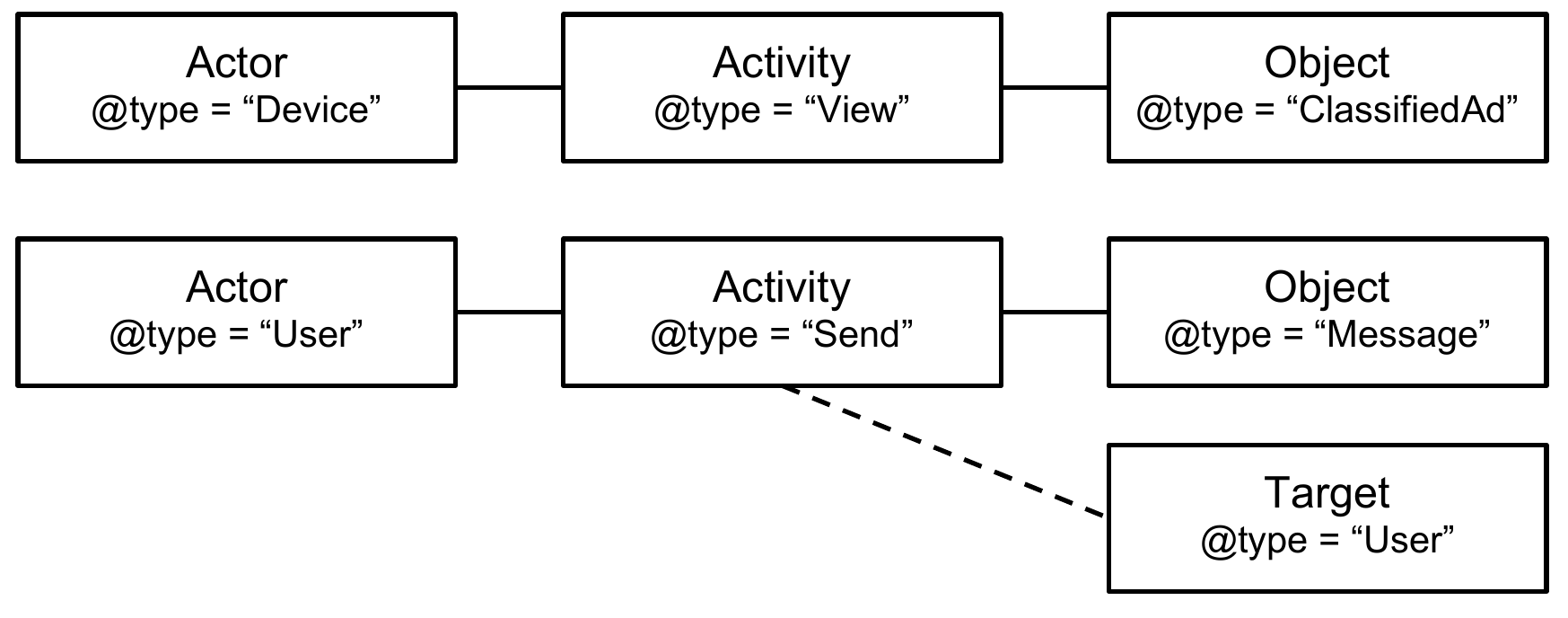}
\caption{Instantiation of the \textit{Actor-Activity-Object} model for View Ad (top) and Send Message (bottom).}
\label{fig:aao-model}
\end{figure}

We have drawn inspiration from Activity Streams 2.0 \cite{activitystreams:online} and Schema.org \cite{schema:online}. In addition to property naming, we also adopt Activity Stream's \textit{Actor-Activity-Object} model to organize user behavior events where a user (Actor) performed an action (Activity) on an item (Object). The convention for an event schema title uses the "Activity Object" format, e.g. "View Item", "Send Message". For an object schema, its title is just the Object name. An example of this model can be seen in Figure \ref{fig:aao-model}. For logged-in events, the actor will be a \textit{User} and will have a unique identifier that corresponds to their account ID. In the case of non logged-in events, we represent the actor type as a \textit{Device}, and the identifier corresponds to the cookie on the device.

We leverage the JSON Schema grammar to organize the schema definition. A schema file is required to contain the following sections:
\begin{itemize}
\item \textit{id}: A URL-format string pointing to the corresponding file hosted on the schema server. Both schema title and version are included to make it unique.
\item \textit{title}: The title of the schema.
\item \textit{properties}: A list of the fields defined in the schema. See Table \ref{tab:event} for an example of the main properties of a user behavior event.
\item \textit{allOf} (optional): The id of the parent schema, from which all properties are inherited by this schema.
\item \textit{required} (optional): An array containing all the mandatory properties of this schema.
\end{itemize}

In the \textit{properties} section, each property definition consists of four parts:
\begin{itemize}
\item \textit{name}: Name of the property.
\item \textit{type}: A primitive JSON type (number, string, enumeration or array) or a reference to an object schema. 
\item \textit{description}: Freeform text to describe the semantic meaning of the property. It can be omitted if the meaning is clear and obvious from its name.
\item \textit{pattern} (optional): A regular expression that the value of the property must match.
\end{itemize}

\begin{table}
\centering
\caption{Main properties of a user behavior event}
\begin{tabular}{|p{1.5cm}|p{6cm}|} \hline
\textbf{Property} & \textbf{Description}\\ \hline
schema & Self-declared schema ID that this event should comply to.\\ \hline
@id & A globally unique id of the event in the universally unique identifier v4 format.\\ \hline
@type & Activity type of the event.\\ \hline
actor & The entity that carried out the activity. It can be a user, a cookie, a business unit or a system component.\\ \hline
object & The object of the activity. Note that the definition here is different from the generic Object in Activity Streams.\\ \hline
intent & Actor's intention of the activity. This property is often used in the interaction events with UI elements. For example, clicking the "Buy" button indicates the intent to purchase the target item.\\ \hline
target & The target of the activity. This property is used in combination with the \textit{intent} property. Its difference from the \textit{object} property can be subtle sometimes.\\ \hline
published & Time stamp of the event.\\ \hline
\end{tabular}
\label{tab:event}
\end{table}

\subsection{Inheritance and reference}
\label{sec:inherit}

\begin{figure*}
\centering
\includegraphics[width=0.9\textwidth]{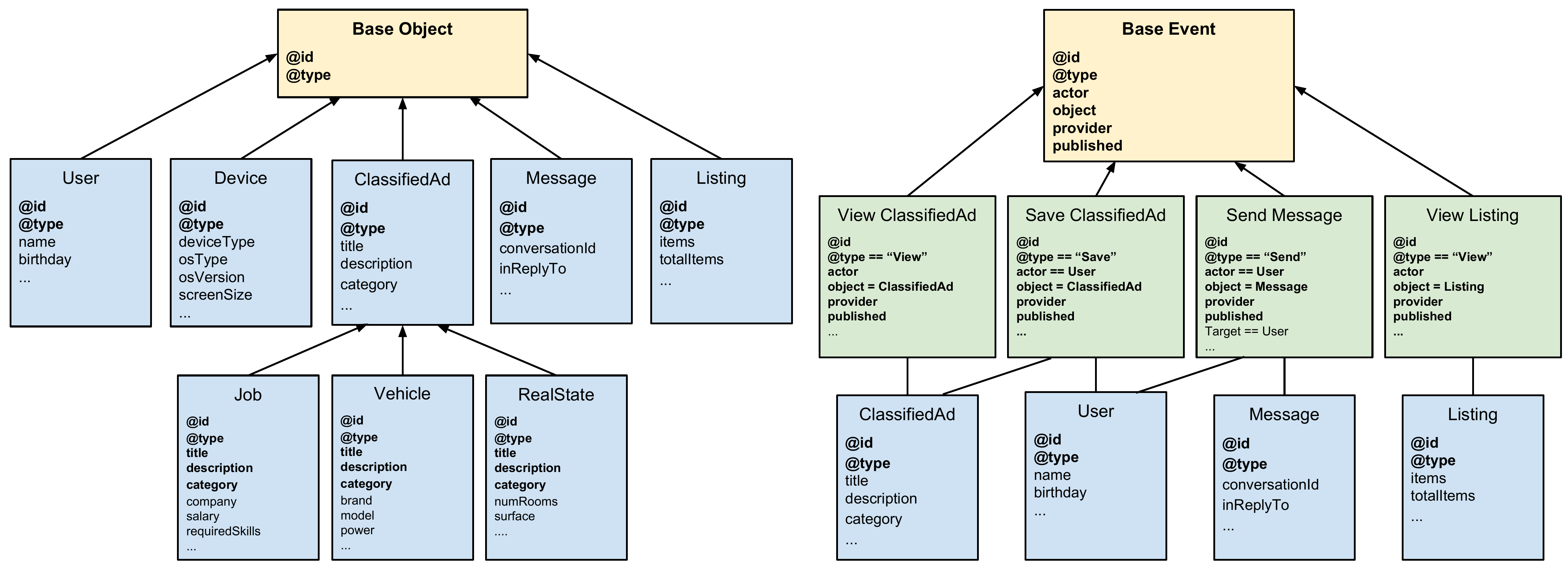}
\caption{Schema inheritance representation for content objects (left) and for events (right). The right figure also includes references to other object schemas. The properties in bold are inherited.}
\label{fig:tree}
\end{figure*}

Our data taxonomy can be represented as a tree as shown in Figure \ref{fig:tree} where the root is a base event schema with the bare minimum requirements to make the data routable in the data platform. All other schemas inherit the properties from the base schema with more properties and restrictions added. 
We use compound properties to group related properties together in hierarchy. When a subset of the properties can be reused by another schema, we define them as a separated object to improve schema reuse.  Similar to events, there is also a base object schema, and all objects such as \textit{ClassifiedAd}, \textit{Message} or \textit{Vehicle} are derived from it. 

The schema definition is rather strict, because we want to minimize the ambiguity on the meaning of the data. At the same time, this does reduce the flexibility of what tenants can send through the pipeline. If we simply allow additional properties in a schema, we will run into situations that two tenants want to send different types or ranges of values using the same property name. This will cause exceptions later in the batch jobs for data serialization or data quality issues in the downstream applications. To support the experimental needs of different tenants without causing schema conflict, we add a special \textit{custom} property to support any additional data they want to send without submitting it for schema review. The idea is similar to chunk folding \cite{Aulbach:mapping} for schema mapping in a multi-tenant database. Chunk folding divides database tables into generic chunks and application-specific ones. We maintain both global and partition-by-tenant datasets. The global datasets only contain defined schema supported by all tenants and simply drop all data in the \textit{custom} property. All custom data must be subfields under the \textit{custom} property and kept in the partition-by-tenant datasets. If the data passes the experiment stage to be used in production, the tenant will go through the schema review and move it out of \textit{custom} to become part of the official schema. Version transformation is not supported for \textit{custom} properties, so we require them to always be string type to avoid potential type conflict in partition-by-tenant batch jobs.

\begin{figure*}
\centering
\includegraphics[width=0.875\linewidth]{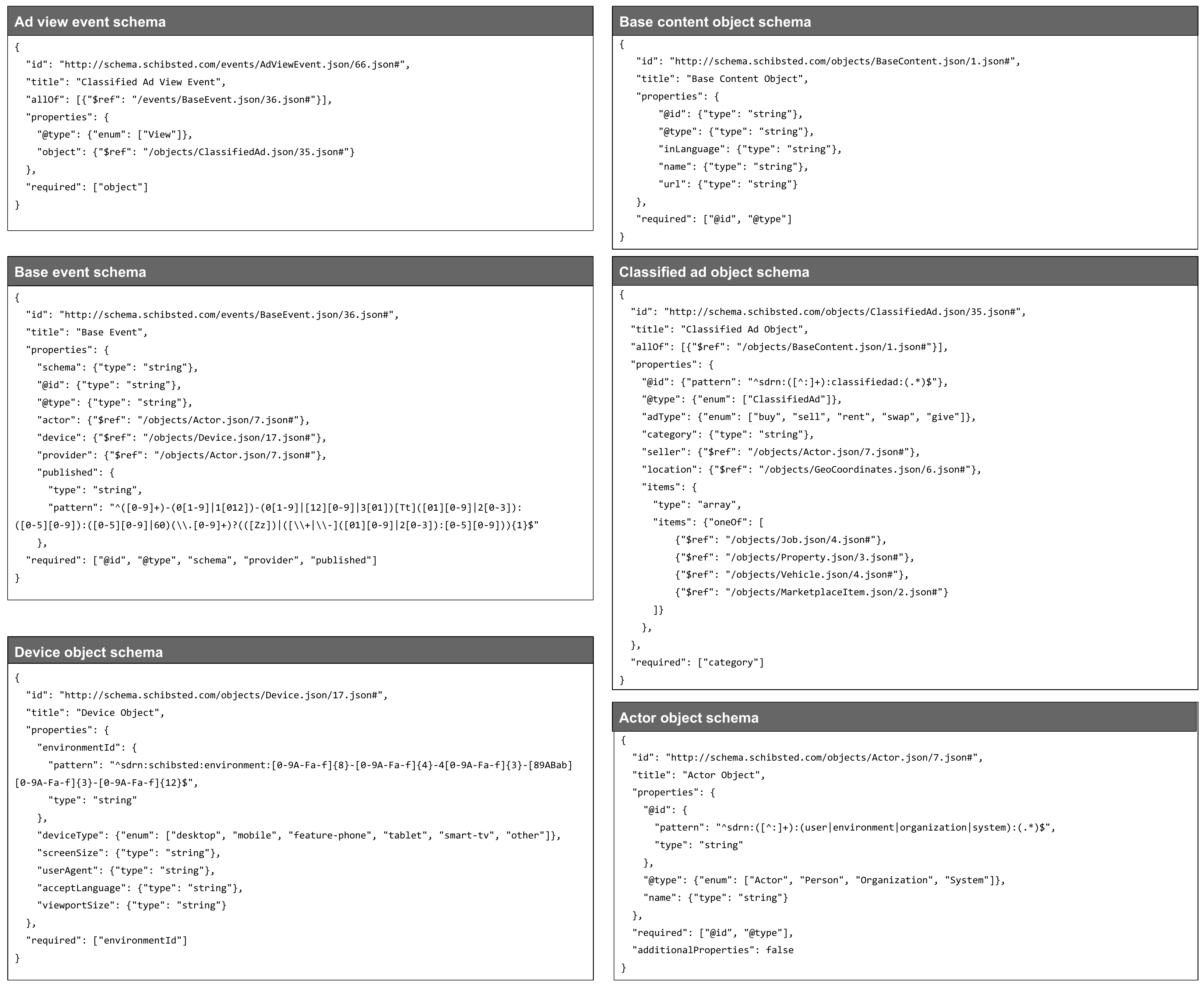}
\caption{Marketplace item view event schema}
\label{fig:schema}
\end{figure*}

\begin{figure*}
\centering
\includegraphics[width=0.875\linewidth]{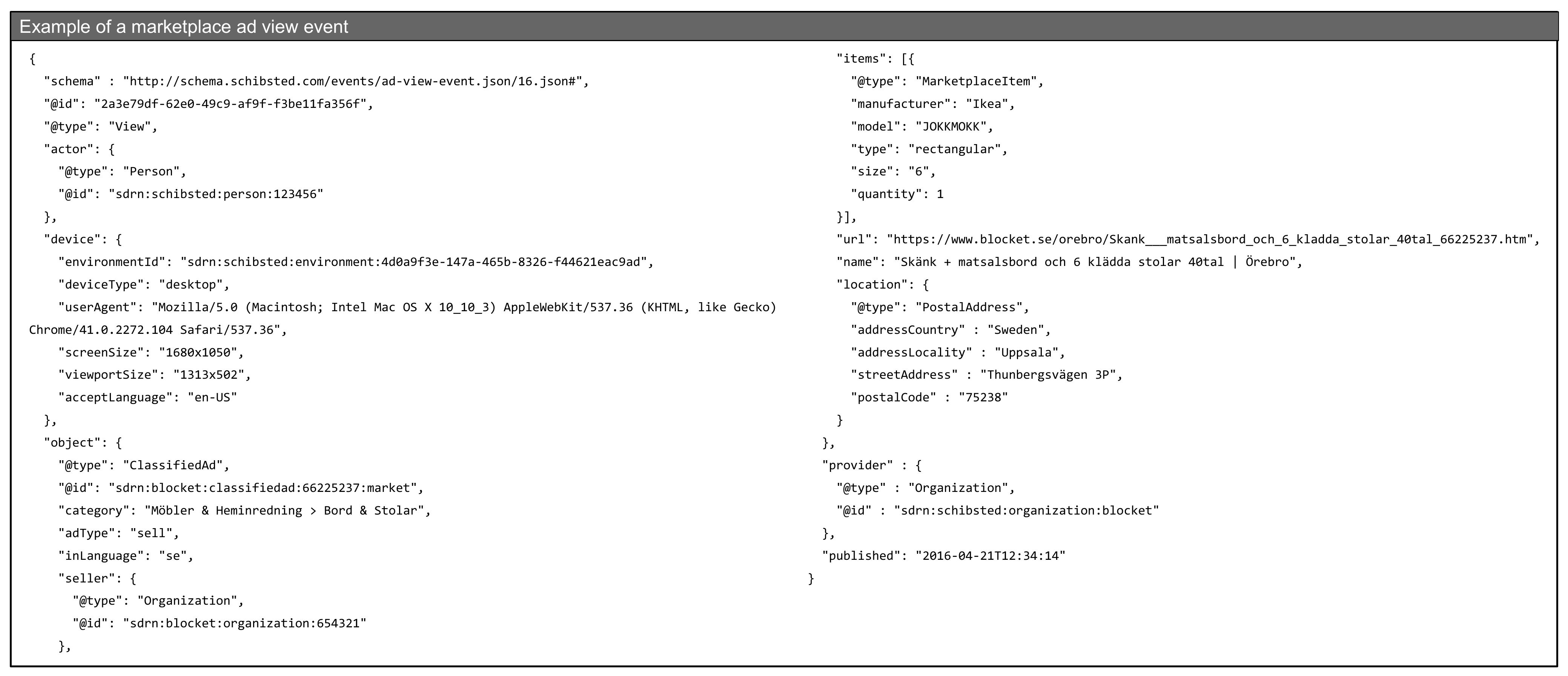}
\caption{An example of a marketplace item view event}
\label{fig:example}
\end{figure*}

We include an example of item view event with related schemas and one sample event data in JSON format in Figure \ref{fig:schema} and \ref{fig:example}. The files are simplified to focus on the features presented in this paper.


\subsection{Update and review process}
\label{sec:process}

The first version of our schema was defined as a joint effort by the event tracker team and the search team who was building a global search service using the data collected by the tracker. Even without all the tooling described in Section \ref{sec:tooling} in place during the early days, this very first version 
enabled training a relevance-based search engine in the new multi-tenant system. Since then, the schema has been growing as more data needs rapidly arose, and more teams joined the efforts. However, we encountered a period with frequent incidents due to schema related data quality issues. Thus we decided to develop a rigorous process and invest in the tooling, so that the schema usage would become more effective and scalable.

Organizationally, schema definition is a data consumer driven process. This is, the team that needs to consume a new type of event will be responsible for defining its schema. The process is overseen by a schema steering committee consisting of representatives from both data producers and consumers to decide on important system design issues. The steering committee members are from different teams - back-end engineering, front-end engineering, data science, product analytics, etc., with long experience in data modeling and processing to provide guidance on defining a schema. The steering committee together with all contributors and reviewers make an active community working on the schema in a collaborative way across the company. The community members change over time as new projects come up and bring in new perspectives. Attracting new members into the community and sharing the knowledge are critical to make this process scalable, not creating bottlenecks in the end-to-end data solution. Documentation as well as schema bootcamps are the main channels for growing the community. Today, the schema repository enjoys a total of 80 contributors across Schibsted and the community keeps growing.

The process of adding a new schema is as follows. When a team needs to consume a certain event that is not tracked yet by a marketplace, they will make a proposal about the new schema in the form of a pull request to the schema repository and invite reviews with the data producer. When a team needs to change the way a certain event is tracked by a marketplace, the process of updating an existing schema is similar, except that they must also invite other consumers of the impacted data to review the pull request. When it is a major change with broad impact to multiple schemas, one or more schema committee members will be part of the review. The latest schemas are available as a Java library, while the latest and previous versions of all the schemas are also hosted at a web server. Approved changes are merged to the master branch, and then released to the schema server and updated in the Java library. To make it easier to understand the schema, the repository also includes example events which are covered by tests to ensure they comply to the latest schemas. 

\section{Schema tooling}
\label{sec:tooling}

\begin{figure}
\centering
\includegraphics[width=\linewidth]{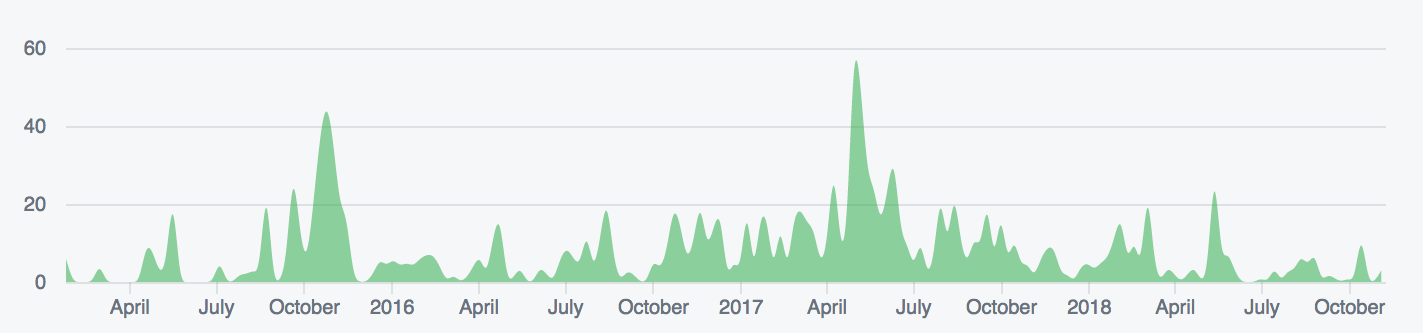}
\caption{Schema evolution history. Contributions to the schema repository in Feb 2015 - Nov 2018.}
\label{fig:history}
\end{figure}

Schema tooling contains the core functionality to support schema evolution - adding a new schema, updating an existing schema, retiring an outdated schema, and enabling data processing across different schema versions. The architectural design of the tooling is highly dependent on the evolution pattern of the schema. We observed frequent changes at the beginning of the project and made a bold assumption that the schema would keep evolving as new features arise with new data needs. Based on the commit history of our schema repository shown in Figure \ref{fig:history}, the assumption is likely to be true. A major revision can temporarily reduce the frequency of changes for a period, but the needs for new changes will always show up later. The schema evolution may slow down as a product becomes mature, but that is not the case for Schibsted when we plan to actively innovate with our marketplaces. Thus we decide to support small and frequent schema changes to empower rapid innovation, and designed the tooling to evolve the schema in a pay-as-you-go model \cite{furche:visionary}, making incremental changes when needed.

In this section, we will first describe the versioning mechanism in Section \ref{sec:versioning}, and then cover how we build a series of test and transformation on top in Section \ref{sec:test} and \ref{sec:transform}. 

\subsection{Schema versioning}
\label{sec:versioning}

For backward compatibility with historical data, we need a versioning system to keep track of both the latest and the previous schemas. To suit the needs of frequent changes, we designed a dual versioning system with a combination of linear and semantic versions.

Linear version is an integer number used to manage the version of a single schema. The life cycle of a schema consists of three stages: created, alive and tomb-stoned. A schema is marked as version 0 upon creation, and each change automatically bumps up one version. When a schema ceases to be of use, it is tomb-stoned by a special pull request that removes all its properties. 
Tomb-stoning pull request is treated in the same way as a normal change, which means it bumps a new version to a schema that does not have any property instead of deleting the schema. In this way, one can easily revive a schema by adding properties back. Most schema definitions have a linear version of 10+ after the system is in production for three years. Some of the most revised ones have 100+ versions now.

In addition to the linear version on each schema, a semantic version of the format MAJOR.MINOR.PATCH is tagged on the whole schema repository. This version is linked to the event tracker explained in Section \ref{sec:tracker}. It marks a snapshot of the latest schema supported by a specific tracker version. A new release of the tracker will trigger a PATCH version bump by default, a MINOR version bump when there are breaking schema changes from the previous release time, or A MAJOR version bump only when the new event tracker does not have backward compatibility of the previous schema. Because the tracker is designed to be schema agnostic, MAJOR bumps happen very rarely. The only two MAJOR bumps we had were due to identifier format change in the tracker.

We intentionally exclude changes in the \textit{custom} property from the versioning system, so that they do not trigger any schema version bump.

\subsection{Schema testing}
\label{sec:test}

We have developed various tests to prevent problematic schema updates. The most basic one is JSON validity test that automatically verifies if the changed schema files violate any JSON grammar. The greater challenge is data provenance. For each schema update, the data consumer driving the change must find out who produces the data and who else also consume the data so as to align with them. To facilitate this process, we want to test two things: 1) whether data producers need to update their ingestion code to comply to the new schema, and 2) whether the new schema will require changes from other data consumers. We build a schema validation tool for the former and change impact testing for the latter.

\textbf{Schema validation} is done by comparing an event to the schema it should comply to. The self-declared \textit{schema} property in the event is selected by default, but users can also specify a schema to use or force to use the latest version. The validator will return the mismatching properties, if the event does not fully comply. The validation test is used in three different ways: as a unit or integration test, as a command line tool, and as an API endpoint. The schema repository uses it as a unit test to check if the proposed change requires updating the examples. Data producers include it in the integration test to make sure that their event trackers produce data according to the correct schema. Data consumers often use it from the command line to debug when their machine learning job fails to load the data correctly. In the data quality checks described in Section \ref{sec:quality}, the schema compliance check samples 1\% from the ingestion stream and send them to the validation endpoint to check if they comply to the required schema.

\textbf{Change impact test} aims to find out if any other consumers will be impacted by the proposed change. Each consumer is asked to upload valid samples of the properties they use. The test will fill the other properties with random valid values based on the new schema to try to generate valid events. If a consumer's sample data is no longer valid, their change impact test will fail and block the schema update pull request. The pull request owner must contact the owner of the failed test to resolve it before they can merge the schema update. The change impact test functions well with schema with strict requirements. However, data consumers sometimes define the schema too loose. For example, they might define a plain string type for an email property and end up seeing unexpected strings later. To overcome this problem, we experiment the opposite "expected failure test" - data consumers anticipate the common data quality issues they might encounter and provide samples that would cause the issues. Those samples should always be invalid according to the schema. If a schema change makes them valid, that means the change is too loose and should tighten its property type and pattern requirements.

\subsection{Schema transform}
\label{sec:transform}

There are four basic operations of schema evolution:
\begin{itemize}
\item \textbf{Add} a new property or a new schema
\item \textbf{Modify} the definition of an existing property
\item \textbf{Remove} an existing property or an existing schema
\item \textbf{Rename} a property or a schema without altering its definition
\end{itemize}

Except \textbf{add}, all the other operations will result in breaking changes for data consumers who are using the changed properties. Based on an analysis about the merged pull requests in our schema repository between April 2015 and March 2017, 64\% of the pull requests contain schema changes other than \textbf{add}. The continuous evolution of schema necessitates data wrangling to cope with frequent breaking changes. Thus we introduced schema transforms. A transform takes the input data in one version and outputs it in another version. Pull requests including breaking changes are also required to submit a transform script indicating how to convert the data from the old schema to the new one. The transforms can go both backward and forward in theory, but we have not see a strong use case in our company for backward transform. In practice, most schema changes are either bug fixes or improvements, so backward transforms for such changes are meaningless and error-prone. For example, removing a property is not backward-compatible, unless we can calculate its value using other properties or set a default value for it. In our production system, we only enforce the forward transform to bring data to the latest schema.

Schema transform may appear similar to the traditional ETL (Extract, Transform and Load) \cite{vassiliadis:etl}. There are indeed some overlaps in the sense that they keep the data clean for data consumers, but schema transform in our solution works directly from the ingestion end and minimizes the variations data consumers must handle. 


\section{Data Quality Tooling (DQT)}
\label{sec:quality}
In order to validate that the data sent by the producers is correct and can be used by the different consumers, we build data quality tooling (DQT) that monitors the data quality end-to-end.


In this section, we first introduce a query language we designed to facilitate data processing, and then describe the DQT usage in the event tracker on the producer end and the DQT checks at the consumer end.

\subsection{JSLT - JSON query and transformation language}
\label{sec:jslt}

The schema transform described in Section \ref{sec:transform} uses Spark \cite{spark:online} in the back-end for performance, but most data consumers working on the schema design are not familiar with Spark. To reduce the learning curve of working with the schema, we developed JSLT, a query and transformation language for JSON to support routing and transformation of data in the data pipeline as well as data quality checks described in Section \ref{sec:DQT}. Thanks to the JSLT language, teams across Schibsted are able to configure their own data flow and data checks without in-depth knowledge of the data pipeline code. Moreover, JSLT can also be used for writing data conversion scripts for external schemas not managed by our system.

The query part is very similar to other JSON languages, supporting object key traversal, array indexing and slicing, boolean and arithmetic operations, function calls, and so on. The transformation part works by writing the fixed formatting in JSON syntax, then using the query part to dynamically insert computed values where desired. Array and object comprehensions, if/else, variable assignments, and user-defined functions are also supported. Objects can also be matched against a template, so that only some keys are redefined or removed while the rest of the object is kept intact. Here is an example of a transform in JSLT that renames and modifies several properties in an event:

\begin{lstlisting}[language=json,firstnumber=1]
{
    "time": round(parse-time(.published, "yyyy-MM-dd'T'HH:mm:ssX") * 1000),
    "device_manufacturer": .device.manufacturer,
    "device_model": .device.model,
    "language": .device.acceptLanguage,
    "os_name": .device.osType,
    "os_version": .device.osVersion,
    "platform": .device.platformType,
    "user_properties": {
        "is_logged_in" : boolean(.actor."spt:userId")
    }
}
\end{lstlisting}

In the example above, 
the \textit{time} field is populated by parsing the timestamp string in the incoming \textit{published} field to seconds since epoch, then multiplying that with 1000 to get milliseconds. The next lines simply move data from the named fields to new fields. At the end, the developer writes a nested object in the output. The \textit{is\_logged\_in} property is true if the input event contains a user ID, false otherwise.

The language is implemented in pure Java and was released as an open-source project in May 2018 \cite{jslt}. Today it performs around 15 billion transforms per day in our data pipeline and has attracted adoption outside Schibsted too.

\subsection{DQT for data producers}
\label{sec:tracker}

An in-house event tracker is provided to data producers as an SDK for different platforms - web, Android, iOS - to embed in their front-end products. Those products are often downloaded to users' mobile devices and have a longer release cycle. While we want to ensure that they send the data using the latest schema, it is beneficial to decouple them from the frequent schema changes as much as possible. 

To achieve that, the schema library is included as a dependency when installing the tracker. It provides event constructor classes and schema validation tests to help data producers send out high quality data. For web applications, they can always load the latest tracker and schema library from the content distribution network. However, for mobile applications, once they are installed on users' devices, the schema library cannot be changed until users update their applications. Hence it is very common to see multiple schema versions coexist in one dataset from mobile platforms. We count on schema transforms to update them to the latest version before storing on disk or passing it to data consumers. Currently we are experimenting with a solution that separates the producer and consumer schemas and always uses schema transforms to connect them. Comparing to the structured and enriched consumer schema, the producer schema is flat, concise and stable. Consumer schema changes such as property renaming and reformatting will not result in a producer schema change.

\subsection{DQT checks for data consumers}
\label{sec:DQT}
Different machine learning teams have different data needs. Using the DQT checks, data consumers can specify detailed checks on incoming data and communicate their quality requirements to data producers. DQT checks are written either in Scala or in JSLT described in Section \ref{sec:jslt}, which allows the requirements to be expressed compactly and safely. In practice, nearly all checks are implemented in JSLT, since most can be defined in a single line. Checks are grouped into modules, where each module is written by one data consumer. Checks consist of meta data, a filter, and the check itself, as shown in the example below:

\begin{lstlisting}[language=json,firstnumber=1]
"user_id_format": {
    "description": "User ID must be a number",
    "solutionUrl": "https://confluence.schibsted.io/...",
    "filter": ".actor.\"spt:userId\"",
    "check" : "test(.actor.\"spt:userId\", \"^sdrn:[^:]+:user:\")"
},
\end{lstlisting}

\begin{figure}
\centering
\includegraphics[width=0.9\linewidth]{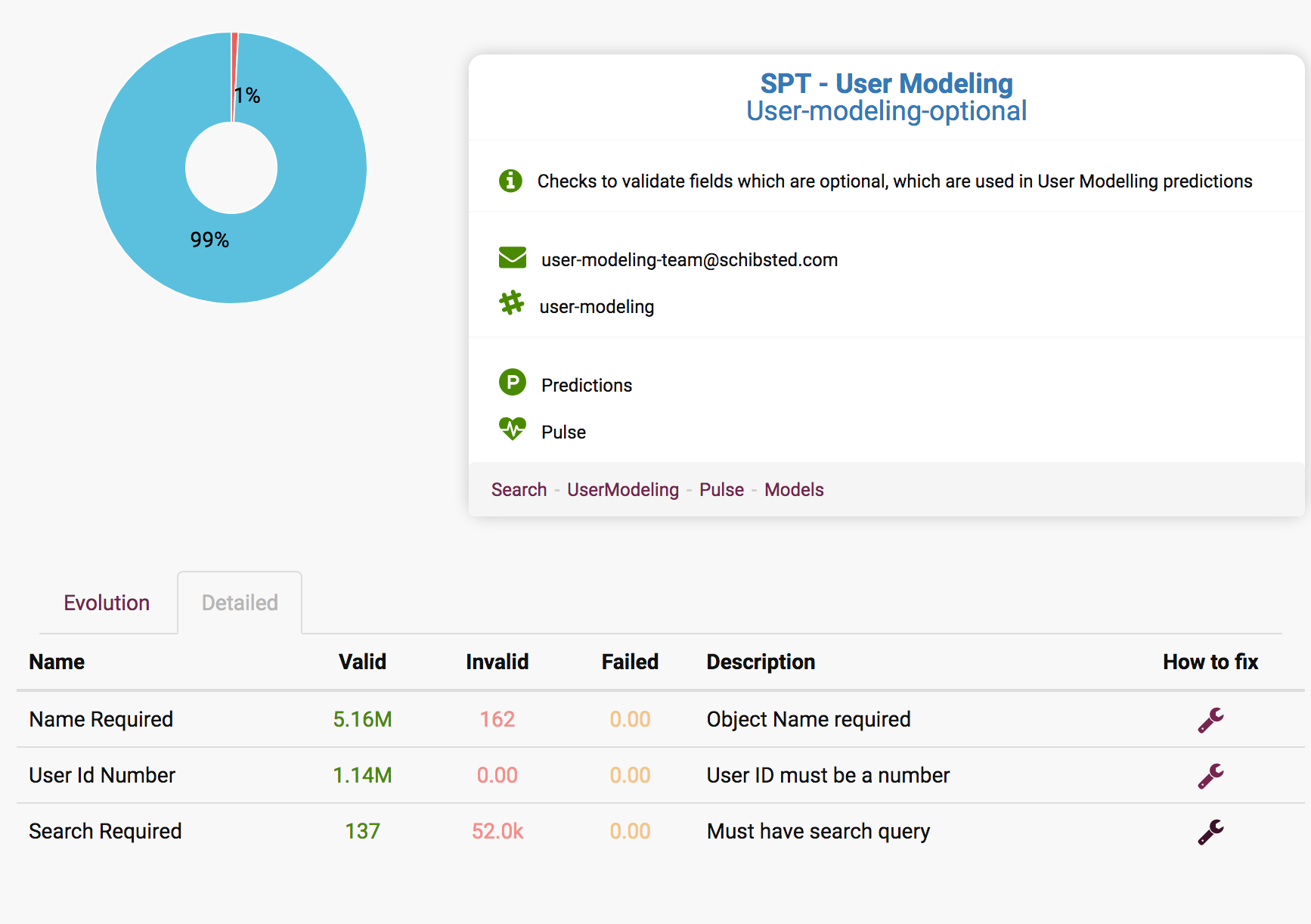}
\caption{DQT dashboard for one tenant's check module}
\label{fig:dqt}
\end{figure}

The \textit{check} is written by the \emph{User Modeling} team to verify that the user ID must comply to the format of "sdrn:<>:user:<>", which is used for Schibsted user IDs. The schema does not always require this, because it handles external account IDs as well. However, the User Modeling team's tool chain does require this to make sure they only process Schibsted users' data, so any tenant who wishes to use User Modeling's services must provide data that matches this check. The \textit{filter} is used to evaluate whether the check applies to a given event or not. In this case, it tests whether the event has a user ID. If it does not (for example because the user is not logged in), the test is not applicable. For events that pass the filter, the check is evaluated. If it returns true (the ID is a number), the metric \textit{user\_id\_number.valid} is incremented. If false, \textit{user\_id\_number.invalid} is incremented. In this way, it is possible to evaluate the percentage of valid events. An example of real results is shown in Figure \ref{fig:dqt}. For querying purpose, the metrics are tagged with event type, the type of event tracker that produced it, and which tenant the event comes from.

In order to give live feedback on data quality, 1\% of the data is sampled from a real-time streaming service for quality checks, and the metrics are streamed into Datadog (\url{https://www.datadoghq.com}), our monitoring service. Users can access dashboards showing the metrics through an online application with various alerting support.

Among the data quality issues consumers encountered before introducing DQT checks, about 80\% are caused by inconsistency due to loose schema definition. DQT checks can detect the ambiguity early and help consumers to tighten the schema to eliminate the caveat. DQT checks can also enforce data quality requirements beyond schema validation. For example, the \textit{age} property is only required to be of the integer type in the schema, but the data consumer might also require it to be within a reasonable range. Such requirements can be difficult or unnecessary to express in the schema. In a multi-tenant environment like Schibsted, the schema is meant to cover all use cases, while specific consumers can have more strict needs than the others.

\section{Conclusion}
\label{sec:conclusion}

In this paper, we present a data quality management system that is based on a semantic schema system that describes user behaviors in marketplaces and supports continuous schema evolution in a multi-tenant system. Moreover, this system is instrumented with the tooling to monitor data quality in the data tracking and processing components in an end-to-end pipeline. The system can be used for managing a fast-changing schema in the single application model as well, but its advantages manifest best in the multi-tenant model where cross-team collaboration is essential but expensive. We believe that many of our learnings can be applicable to companies that are undergoing a transformation from product silos to the multi-tenant model as Schibsted, and for companies that are just starting to build a new data-driven organization and are currently establishing its basis.

For future work, we plan to integrate privacy compliance checks, which are currently implemented in a separated system. By specifying in the schema if a property may contain personal information, we will be able to automatically detect if a dataset contains personal data and raise an alert if the dataset is not stored with the correct security and privacy compliance settings. We also plan to explore auto generation of schema transforms to simplify the process of making schema changes.


\bibliographystyle{ACM-Reference-Format}
\bibliography{schema}

\balance

\end{document}